\documentclass[sigconf]{acmart}

\usepackage{xspace}
\usepackage{textcomp}
\usepackage{soul}
\usepackage{marginnote}
\usepackage{hyperref}
\usepackage{amsfonts}
\usepackage{graphicx}
\usepackage{float}
\usepackage{subfig,rotate}
\usepackage{verbatim}
\usepackage{comment}
\usepackage{versions}
\usepackage{multirow}
\usepackage{longtable}
\usepackage{booktabs}
\usepackage{makecell}
\usepackage{caption}
\usepackage{acro}
\usepackage{tabularx}
\usepackage{adjustbox}
\usepackage{listings}
\usepackage{color}

\def\bitcoin{%
  \leavevmode
  \vtop{\offinterlineskip %\bfseries
    \setbox0=\hbox{B}%
    \setbox2=\hbox to\wd0{\hfil\hskip-.03em
    \vrule height .3ex width .15ex\hskip .08em
    \vrule height .3ex width .15ex\hfil}
    \vbox{\copy2\box0}\box2}}

\definecolor{dkgreen}{rgb}{0,0.6,0}
\definecolor{gray}{rgb}{0.5,0.5,0.5}
\definecolor{mauve}{rgb}{0.58,0,0.82}

\acmConference[Technical Report]{}{July}{2019}
\setcopyright{none}
\renewcommand\footnotetextcopyrightpermission[1]{}

\begin{document}

\title[BlockTag: Design and applications of a tagging system for blockchain analysis]{\huge BlockTag: Design and applications of a tagging system for blockchain analysis}

\author{Yazan Boshmaf}
\affiliation{
  \institution{Qatar Computing Research Institute, HBKU}
%  \city{Doha}
%  \country{Qatar}
}
\author{Husam Al Jawaheri}
\affiliation{
  \institution{Qatar University}
%  \city{Doha}
% \country{Qatar}
}
\author{Mashael Al Sabah}
\affiliation{
  \institution{Qatar Computing Research Institute, HBKU}
%  \city{Doha}
%  \country{Qatar}
}
\renewcommand{\shortauthors}{Boshmaf et al.}

\begin{abstract}
Annotating blockchains with auxiliary data is useful for many applications. For example, e-crime investigations of illegal Tor hidden services, such as Silk Road, often involve linking Bitcoin addresses, from which money is sent or received, to user accounts and related online activities. We present BlockTag, an open-source tagging system for blockchains that facilitates such tasks. We describe BlockTag's design and present three analyses that illustrate its capabilities in the context of privacy research and law enforcement.
\end{abstract}

%TODO: replace this section with code generated by the tool at https://dl.acm.org/ccs.cfm
% <concept_id>10002978.10002991.10002994</concept_id>
% <concept_desc>Security and privacy~Pseudonymity, anonymity and untraceability</concept_desc>
% \ccsdesc[500]{Security and privacy}

% -- end of section to replace with generated code

%\keywords{Blockchain, Tagging, Bitcoin, Privacy, Law Enforcement} % TODO: replace with your keywords

\maketitle

\section{Introduction}
\label{sec:introduction}

Public blockchains contain valuable data describing financial transactions. For example, Bitcoin's raw blockchain data alone is 160 GB as of March 2018, and is growing rapidly. This data holds the key to understanding different aspects of cryptocurrencies, including their privacy and market dynamics. Blockchain analysis systems, such as BlockSci~\cite{kalodner2017blocksci}, have enabled blockchain science by addressing three pain points, namely poor performance, limited capabilities, and a cumbersome programming interface.

\paragraph{Overview}

We present BlockTag, a tagging system for blockchains. BlockTag uses vertical crawlers to automatically annotate blockchain data with tags, mappings between block, transaction, or address identifiers and auxiliary data describing the tagged identifiers. For example, the system can tag Bitcoin address with the Twitter user account of its owner. BlockTag also provides a novel blockchain query interface with advanced capabilities, such as clustering, linking and searching, which are important for privacy research and law enforcement. For example, BlockTag provides best-effort answers to high-level queries in Bitcoin e-crime investigations, such as ``which Twitter user accounts paid $\geq\bitcoin10.0$ to Silk Road in 2014.''

\begin{figure}
\centering
\includegraphics[width=0.9\linewidth]{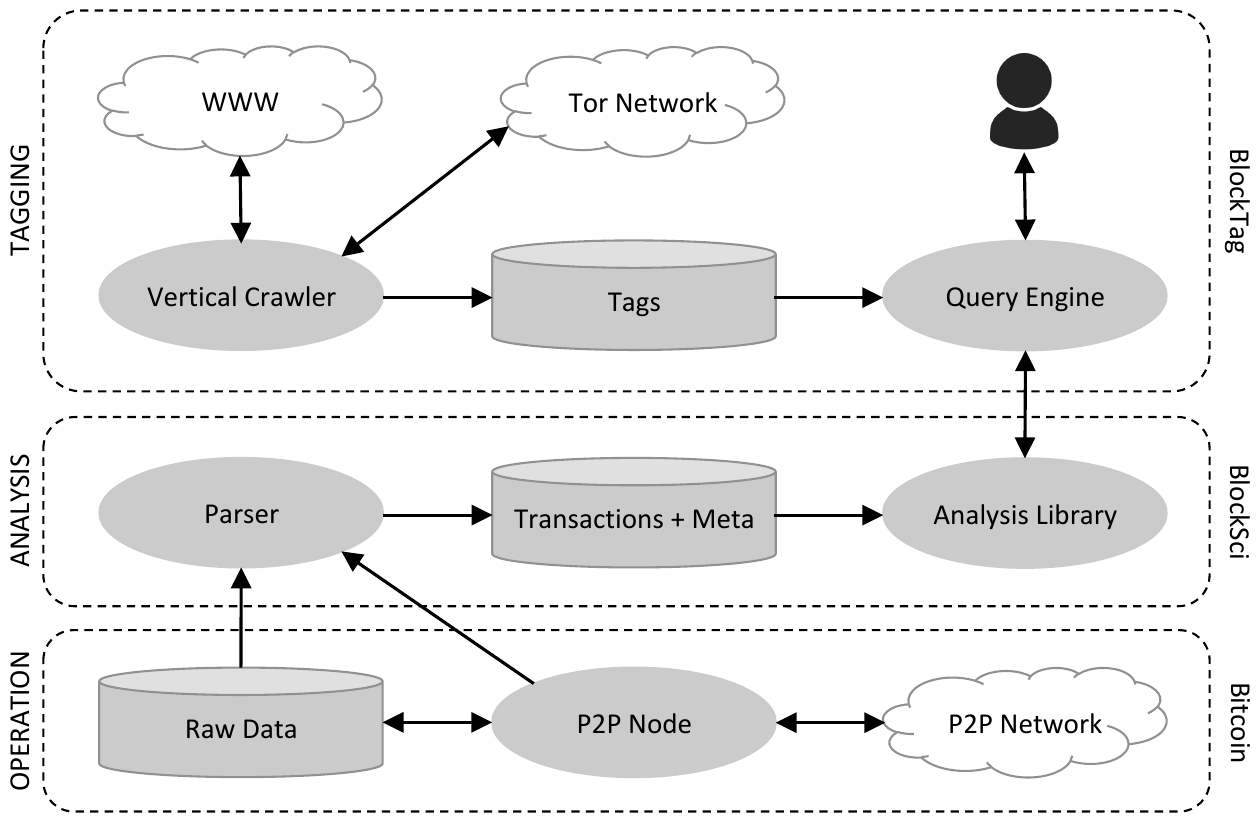}
\caption{Layered blockchain system architecture.}
\label{fig:architecture}
\end{figure}

\paragraph{Design}
We start with the observation that most blockchain analysis systems transform raw blockchain data into a stripped-down, simple data structure that can fit in or map to memory. Therefore, information auxiliary to core transaction data, such as scripts, hashes, or annotations in general, cannot be part of this data structure and must have their own mappings. This naturally leads to a layered system architecture, where a tagging layer sits on top of an analysis layer, with a well-defined and extendable interface between them, as shown in Figure~\ref{fig:architecture}. In our implementation, we chose BlockSci as a blockchain analysis system because it is hundreds times faster than its contenders. Moreover, BlockSci exposes a programming interface in C\texttt{++} to extend its core analysis library, along with a Python wrapper for defining high-level analytical tasks.

BlockTag is shipped with batteries included. First, it implements four vertical crawlers that are configured to annotate Bitcoin addresses with three types of tags: user tags representing BitcoinTalk\footnote{\url{https://bitcointalk.org}} and Twitter user accounts, service tags representing service providers that are indexed by Ahmia\footnote{\url{https://ahmia.fi}} search engine, and text tags representing user-generated textual labels submitted to Blockchain.info.\footnote{\url{https://blockchain.info}} Second, BlockTag is not limited to Bitcoin. The vertical crawlers can be configured to scrape auxiliary data of other cryptocurrencies, including Litecoin, Namecoin, and Zcash. Smart contract platforms, such as Ethereum and EOS, are outside our scope. Third, BlockTag extends BlockSci's analysis library and implements a programming interface that enables analysts to query transactions by their properties, including tags. Fourth, BlockTag allows analysts to manually annotate blockchains with custom tags at the block, transaction, and address level.

Analysts start blockchain investigations using a Jupyter notebook that imports BlockTag's Python package. The package exposes a \texttt{chain} object representing the blockchain. Each block, transaction, and address has a \texttt{tags} object mapping it to some JSON-serializable auxiliary data. Selecting, grouping, and aggregating transactions is straightforward and is provided through a simple query interface.

\paragraph{Deployment}
We deployed BlockTag on a single, private, server-grade machine in January 2018 for about three months. As of March 2018, the crawlers have ingested about 5B tweets, 2.2M BitcoinTalk user profiles, 1.5K Tor onion pages, and 30K Blockchain.info labels. This has resulted in 45K user, 88 service, and 29K text tags.

\paragraph{Applications}
We demonstrate BlockTag's novel capabilities with three applications, focusing on Bitcoin and Tor hidden services. 

{\em1) Linking:} We show it is relatively easy to link users of social networks to Tor hidden services through Bitcoin payments. We were able to link 125 user accounts to 20 service providers, which include illegal and controversial ones, such as Silk Road and The Pirate Bay. While one may expect a better level of privacy when using Bitcoin, we recall that it is pseudo-anonymous by design and lacks retroactive operational security, as described by Satoshi~\cite{nakamoto2008bitcoin}. From a law enforcement perspective, BlockTag offers a valuable capability that is useful in e-crime investigations. In particular, showing a provable link between a user account on a website and illegal activities on the Dark Web can be used to secure a subpoena to collect more information about the user from the website's operator~\cite{marquardt2014dpr}.

{\em2) Market economics:} We analyze the market of Tor hidden services by calculating their ``balance sheets.'' We show that WikiLeaks is the highest receiver of payments in terms of volume, with 26.4K transactions. In terms of total value of incoming payments, however, Silk Road tops the list with more than $\bitcoin29.6$K received on its address. We also observe that the money flowing in and out of service addresses is nearly the same. This suggests that service providers do not keep their Bitcoins on the same address they use for receiving payments, but rather distribute them to other addresses. Third, from the last transaction dates of these addresses, we found that all but three of the top-10 revenue making service providers are active in 2018. This, however, does not mean the three services have stopped making Bitcoin transactions, as the service providers might have used different addresses that the crawlers have not found.

{\em3) Forensics:} We link 24.2K users and 202 labels to MMM, which is one of the world's largest Ponzi scheme. All of these users are BitcoinTalk users who are mostly male, 20--40 years old, and located worldwide in more than 80 countries. We found that only 313 users have made one or more activities and engaged with the forum once a day, on average. After further analysis, it turned out these user accounts were created as part of the ``MMM Extra'' scheme, which promises ``up to 100\% return per month for performing simple daily tasks that take 5--15 min.'' We also used BlockTag to retrieve and model MMM transactions as a graph. This graph consisted of 14.3K addresses and 32.K transactions. We found that two of the top-10 ranked addresses, in terms of their PageRank, have been flagged on BitcoinTalk as scammer addresses. As of March 21, these addresses has received more than $\bitcoin$2M combined.

\section{Background and related work}
\label{sec:related_work}
Research in blockchain and cryptocurrencies has gained a significant momentum over the years~\cite{bonneau2015sok}. In what follows, we present the background and related work and contrast it to ours.

\paragraph{Analysis systems}
Blockchain analysis systems parse and analyze raw transaction data for many applications. Recently, Kalodner et al. proposed BlockSci~\cite{kalodner2017blocksci}, an open-source, scalable blockchain analysis system that supports various blockchains and analysis tasks. BlockSci incorporates an in-memory, analytical database, which makes it several hundred times faster than its contenders. While there is a minimal support for tagging in its programming interface, BlockSci is designed for analysis of core blockchain data. At the cost of performance, annotation and tagging can be integrated into the analysis pipeline through a centralized, transactional database. For example, Spagnuolo et al. proposed BitIodine~\cite{spagnuolo2014bitiodine}, an open-source blockchain analysis system that supports tagging through address labels. However, BitIodine, relies on Neo4j~\cite{miller2013graph}, a general-purpose graph database that is not designed for blockchain data and its append-only nature, which makes it inefficient for common blockchain analysis tasks, such as address linking. In contrast, BlockTag is the first open-source tagging system that fills this role.

\paragraph{Linking}
The impact of Bitcoin address linking on user anonymity and privacy has been known for a while now~\cite{reid2013analysis,jordan2013fistful,dupont2015toward,fleder2015bitcoin}. Fergal and Martin~\cite{reid2013analysis} showed that passive analysis of public Bitcoin information can lead to a serious information leakage. They constructed two graphs representing transactions and users from Bitcoin's blockchain data and annotated the graphs with auxiliary data, such as user accounts from BitcoinTalk and Twitter. The authors used visual content discovery and flow analysis techniques to investigate Bitcoin theft. Alternatively, Fleder et al.~\cite{fleder2015bitcoin} explored the level of anonymity in the Bitcoin network. The authors annotated addresses in the transaction graph with user accounts collected from BitcoinTalk in order to show that users can be linked to transactions through their public Bitcoin addresses. These studies show the value of using public data sources for Bitcoin privacy research and law enforcement, which is our goal behind designing BlockTag.

\paragraph{Tor hidden services and black markets}
Tor hidden services have become a breeding ground for black markets, such as Silk Road and Agora, which offer illicit merchandise and services~\cite{biryukov2014content,moore2016cryptopolitik}. Moore and Rid~\cite{moore2016cryptopolitik} studied how hidden services are used in practice, and noted that Bitcoin was the dominant choice for accepting payments. Although multiple studies~\cite{fleder2015bitcoin,meiklejohn2013fistful} showed that Bitcoin transactions are not as anonymous as previously thought, Bitcoin remains the most popular digital currency on the Dark Web~\cite{castillo2016bitcoin}, and many users choose to use it despite its false sense of anonymity. Recent research explored the intersection between Bitcoin and Tor privacy~\cite{biryukov2014deanonymisation,biryukov2015bitcoin}, and found that legitimate hidden service users and providers are one class of Bitcoin users whose anonymity is particularly important. Moreover, Biryukov et al.~\cite{biryukov2014content} found that hidden services devoted to anonymity, security, human rights, and freedom of speech are as popular as illegal services. While BlockTag makes it possible to link users to such services, we designed it to help analysts understand the privacy threats, identify malicious actors, and enforce the law.

\paragraph{Forensics}
Previous research showed that cryptocurrencies, Bitcoin in particular, have a thriving market for fraudulent services, such as fake mining, wallets, exchanges, and Ponzi schemes~\cite{vasek2015there, bohr2014uses}. Recently, Bartoletti et al.~\cite{bartoletti2018data} proposed a data mining approach to detect Bitcoin addresses that are involved in Ponzi schemes. The authors manually collected and labeled Bitcoin addresses from public data sources, defined a set of features, and trained multiple classifiers using supervised machine learning. The best classifier correctly labelling 31 addresses out of 32 with 1\% false positives. Interestingly, MMM was excluded because it had a complex scheme. In concept, BlockTag complements such techniques by providing an efficient and easy way to collect and explore data that is relevant to the investigation. This data can be then analyzed using different techniques with the help of existing tools~\cite{vasek2018analyzing}.

\section{Design and architecture}
\label{sec:design}
BlockTag is designed for a layered system architecture. As depicted in Figure~\ref{fig:architecture}, each layer in the blockchain stack is responsible for a separate set of tasks and can interact with other layers through programmable interfaces. We present a high-level view of BlockTag's design, and leave the details in the technical report.

\subsection{Tags}
\label{sec:tags}

In BlockTag, a tag is a mapping between a block, a transaction, or an address identifier and a list of JSON-serializable objects. Each object specifies the type, the source, and other information representing auxiliary data describing the tagged identifier. As raw blockchain data is stored in a format that is efficient for validating transactions and ensuring immutability, the data must be parsed and transformed it into a simple data structure that is efficient for analysis. For example, BlockSci uses a memory-mapped data structure to represent core transaction data as a graph. All other transaction data, such as hashes and scripts, are stored separately as mappings that are loaded when needed. BlockTag follows this design choice, and uses a persistent key-value database, such as RocksDB~\cite{rocksdb}, with an in-memory cache in order to store and manage blockchain tags, as they can grow arbitrarily large in size.

BlockTag defines four types of tags, namely user, service, text, and custom tags. A user tag represents a user account on an online social network, such as BitcoinTalk and Twitter. A service tag represents an online service provider, such as Tor hidden services like Silk Road and The Pirate Bay. A text tag represents a user-generated textual label, such as address labels submitted to Blockchain.info. A custom tag can hold arbitrary data, including other tags, and is usually used when creating tags manually by analysts.

In BlockTag, tags are created, updated, and removed at the block, transaction, or the address level. Listing~\ref{listing:create_tag} shows how to create a user tag mapping Bitcoin's genesis address to Satoshi's BitcoinTalk user account. The \texttt{append} flag indicates whether the value defined in this tag should be appended to the existing list, as the address can have other tag values defined already.
\begin{lstlisting}[caption={Creating a tag.},label={listing:create_tag}]
import blocktag
chain = blocktag.Blockchain('/path/to/blockchain/data/')
chain.tag(
	level='address',
	key='1A1zP1eP5QGefi2DMPTfTL5SLmv7DivfNa',
	value=[{
		'type': 'user',
		'source': 'bitcointalk',
		'info': {
			'id': 3,
			'account': 'satoshi',
			'num_posts': 575,
			'num_activities': 364,
			'position': 'founder',
			'date_registered': '2009-11-19 19:12:39',
			'last_seen': '2010-12-13 16:45:41'
		}
	}],
	append=False
)
\end{lstlisting}
A direct, read-only access to tags is possible at any level through the \texttt{tags} object of a block, a transaction, and an address. By default, BlockTag returns the tag of an identifier at a given level along with the tags of identifiers from lower levels. This means it is sufficient to tag only addresses in order to annotate the whole blockchain.

\subsection{Vertical crawlers}
\label{sec:vertical_crawler}

In BlockTag, a vertical crawler is used to scrape a data source, typically an HTML website or a RESTful API, in order to automatically create block, transaction, or addresses tags of a particular type using a website-specific parser. A crawler can be configured to run according to a crontab-like schedule, and to bootstrap on the first run with previously crawled raw HTML/JSON data, which can also be used to initialize blockchain tags. Listing~\ref{listing:user_crawler} shows how to run a BitcoinTalk user crawler at the address level everyday at midnight.
\begin{lstlisting}[caption={Scheduling a crawler.},label={listing:user_crawler}]
chain.crawl(
	level='address',
	config={
		'type': 'user',
		'source': 'bitcointalk',
		'schedule': '0 0 * * *',
		'data': '/path/to/bitcointalk/data/'
	}
)
\end{lstlisting}

BitcoinTalk is one of the most popular Bitcoin forums with more than 2.2M users. In fact, as of July 2018, the forum contained about 42.2M posts, which makes it a good data source to collect public Bitcoin addresses and their associated user accounts. Behind the scene, \texttt{chain.crawl()} in Listing~\ref{listing:user_crawler} spawns a crawler that downloads user account pages through a URL that is unique for each user account. The HTML pages are then parsed to find Bitcoin addresses using regular expressions. As a Bitcoin address is a base58 encoded identifier of 26--35 alphanumeric characters, beginning with the number 1 or 3, the crawler uses the regex \texttt{*[13][a-km-zA-HJ-NP-Z1-9]{25,34}} and eventually creates or updates a user tag for the matched address.

In addition to a BitcoinTalk user crawler, BlockTag implements a Twitter user crawler that consumes Twitter's API, a Tor hidden service crawler that scrapes onion landing pages of Ahmia-indexed service providers, and a Blockchain.info text crawler that scrapes textual labels that are self-signed by address owners or submitted by arbitrary users. By default, the vertical crawlers create Bitcoin address tags, but can be configured to scrape auxiliary data of other cryptocurrencies, including Litecoin, Namecoin, and Zcash.

\subsection{Query engine}
\label{sec:query_engine}

BlockTag query engine is inspired from NoSQL document databases, such as MongoDB~\cite{chodorow2013mongodb}, where queries are specified using a JSON-like structure. Selecting, grouping, and aggregating transactions is provided through a simple query interface.

To write a query, the analyst starts with specifying block, transaction, or address properties to which the results should match using the \texttt{where} parameter. BlockTag treats each property as having an implicit boolean AND. It natively supports boolean OR queries, but the analyst should use a special \texttt{\$or} operator to achieve it. In addition to exact matches, BlockTag has operators for string matching, numerical comparisons, etc. The analyst can also specify the properties by which the results are grouped using the \texttt{group\_by} parameter. Finally, the analyst can specify which properties to return per result with the \texttt{select} parameter. While this query interface is suitable for many tasks, BlockTag's Python package also exposes lower-level functionality to analysts who have tasks with more sophisticated requirements. Listing~\ref{listing:query_transactions} shows how to finds Twitter user accounts who paid $\geq\bitcoin10.0$ to Silk Road in 2014.

\begin{lstlisting}[caption={Querying a blockchain.},label={listing:query_transactions}]
accounts = chain.query(
    level='transaction',
    select= 'input.address.tag.info.account',
    where={
        'input': {
            'address':
                'tag': {
                    'type': 'user',
                    'source': 'twitter'
                }
            }
        },
        'output': {
            'address': {
                'tag': {
                    'type': 'service',
                    'source': 'tor',
                    'info': { 
                        'provider': { '$like': 'silkroad' }
                    }
                }
            }
        },
        'time': '2014'
    },
    group_by='input.address.tag.info.id',
    having='sum(input.value) >= (10.0 * 10**7)',
    clustering={ 'source': 'inputs', 'method': 'original' }
)
\end{lstlisting}

One important capability of BlockTag's query engine is address clustering~\cite{meiklejohn2013fistful}, which can be configured to operate on a particular source, namely inputs, outputs, or both, using one of the supported clustering methods, all through the \texttt{clustering} parameter. Address clustering expands the set of Bitcoin addresses that are mapped to a unique user, service, or text tag through a technique called closure analysis. As a result, this allows the analyst to identify more links between different tags by considering a larger number of transactions in the blockchain.

BlockTag supports multiple address clustering methods. The first method is the \texttt{original} closure heuristic proposed by Meiklejohn et al.~\cite{meiklejohn2013fistful}, which works as follows: If a transaction has addresses $A$ and $B$ as inputs, then $A$ and $B$ belong to the same cluster. The rationale behind this heuristic is that such addresses are highly likely to be controlled by the same entity, as they are signed by the private keys whose owner performed the transaction. While efficient, this method can result in large clusters that include addresses which belong to different entities, due to mixing services and CoinJoin transactions. In order to tackle this issue, BlockTag implements a novel \texttt{minimal} clustering method that prematurely terminates the original clustering method before the clusters grow to their maximum size. Minimal clustering includes a final trimming phase to find clusters that share at least one address and consequently merges them, after which they are removed. Doing so ensures that the clusters are mutually-exclusive and likely to belong to separate entities, but also means the clusters are smaller than usual, reducing the chance of linking different tags as a result.

\section{Real-world deployment}
\label{sec:deployment}

We now describe our experience in deploying BlockTag.

\subsection{Ethical considerations}

BlockTag's functionality depends on tags that map blocks, transactions, and addresses to user accounts, service providers, text labels, and other types of tags. This allows BlockTag to link tags to each other by findings blockchain transactions involving tag identifiers. For example, it is sufficient to show a transaction from Alice's address to Bob's address to link them together. Tag values represent auxiliary data that is collected from public sources, which include social networks, Tor hidden services, and blockchains. As such, we are faced with two privacy-related ethical concerns, namely linking and data collection. In what follows, we discuss the actions we took to address them, as we worked with our institution's IRB board to approve BlockTag deployment.

First of all, the information gathered from anonymous cryptocurrency payments without linking is often limited and non-actionable for privacy research and law enforcement. BlockTag is designed to address this limitation, building on top of previous studies that showed the feasibility, utility, and value of linking users through Bitcoin transactions and public data sources~\cite{reid2011analysis,meiklejohn2013fistful}. BlockTag does not put users at any additional risk, but rather exposes existing ones and corrects common misconceptions, such as Bitcoin being a private or anonymous online payment system. When needed, we reached out to legitimate users whose privacy is at risk, and informed them about how their Bitcoin transactions link to their online activities and what they can do about it. We also posted a notice on BitcoinTalk forum concerning deanonymizating Tor hidden service users.\footnote{\url{https://bitcointalk.org/index.php?topic=2602885}}

Concerning data collection, our deployment uses crawlers which target solely public data sources. The crawlers are polite, passive, and respect \texttt{robots.txt} instructions. This means we do not collect data from sources that require authentication, payment, or email exchange. Also, all collected data is secured and stored on our private infrastructure whose access is restricted to authorized researchers.

Finally, we have shared our deployment plans with a few stakeholders in order to get an early feedback. In response, we engaged with the U.S. Federal Trade Commission, a national financial regulatory authority, two law firms, and an international news agency which were interested in BlockTag and its potential for protecting users, enforcing the law, and uncovering cyber criminals, respectively. This also indicates that evidence acquired through BlockTag is admissible in the court of law.  

\subsection{Setup}

We deployed BlockTag on a single machine from Jan 1 to March 21, 2018. The machine was running Ubuntu v16.04.4 LTS, Bitcoin Core v0.16.0, and BlockSci v0.5.0 on two 2GHz quad-core CPUs, 128GB of system memory, and 2TB of network-attached storage.

We used BlockTag to tag Bitcoin's blockchain at the address level. As of March 2018, the crawlers have ingested nearly 5B tweets, 2.2M BitcoinTalk profiles, 1.5K Tor onion pages, and 30K Blockchian.info labels, resulting in 45K user, 88 service, and 29K text tags. We used a previously collected dataset consisting of 4.8B tweets, which were posted in 2014, to bootstrap Twitter user tags. Moreover, for the first application where we link users to services, we configured address clustering for inputs from user tags using the minimal clustering method. We summarize the created tags in Table~\ref{table:tags_summary}. 

\begin{table}
{\small
\begin{tabular}{llrr}\toprule
& & \multicolumn{2}{c}{\# addresses}\\
\cmidrule(rr){3-4}
Source & Type &  Original & Clustering\\ \midrule
BitcoinTalk & User & 40,970 & 19,213,141\\
Twitter & User & 4,183 & 623,189\\
Tor Network & Service & 88 & --\\
Blockchain.info & Text & 29,643 & --\\
\bottomrule
\end{tabular}
}
\caption{Summary of created tags.}
\label{table:tags_summary}
\end{table}

\section{Applications}
\label{sec:applications}

We demonstrate the capabilities of BlockTag in the following. 

\subsection{Linking users to services}
\label{sec:deanonymization}

In e-crime investigations of illegal Tor hidden services, such as Silk Road, analysts often try to link cryptocurrency transactions to user accounts and activities. This can start with a known transaction that is part of a crime, such as a Bitcoin payment to buy drugs on Silk Road. Instead, a wider search criteria can be used to understand the landscape of activities of illegal services, such as finding service providers that receive the most payments. Either way, the analysts need to link users to services. In BlockTag, this can be achieved in a single query, as shown in Listing~\ref{listing:linking}.

\begin{lstlisting}[caption={Linking different tags via transactions.},label={listing:linking}]
user_service_txes = chain.query(
    level='transaction',
    select= ['input.address.tag.info.account', 'output.address.tag.info.provider', 'self.txes'],
    where={
        'input': {
            'address': {
                'tag': { 'type': 'user' }
            }
        },
        'output': {
            'address': {
                'tag': {
                    'type': 'service',
                    'source': 'tor'
                }
            }
        }
     },
    group_by=['input.address.tag.info.id', 'output.address.tag.info.id'],
    clustering={ 'source': 'inputs', 'method': 'minimal' }
)
\end{lstlisting}

This resulted in linking 28 Twitter user accounts to 14 service providers via 167 transactions and 97 BitcoinTalk user accounts to 20 service providers via 115 transactions. Some of these users were linked to multiple service providers. In total, 125 users were linked to 20 services. The results suggest that although Twitter users are smaller in number compared to BitcoinTalk users, they are more active and have a larger number of transactions with services. In fact, some of these users are ``returning customers,'' as they have performed multiple transactions with the same service provider.

\begin{table}
\centering
{\small
\begin{tabular}{l c c c} \toprule
 & \multicolumn{3}{c}{\# linked users}\\
\cmidrule(lr){2-4}
Name & Twitter & BitcoinTalk & Total\\ \midrule
WikiLeaks & 11& 35 & 46\\
Silk Road & 4 & 18 & 22\\
Internet Archives & 3 & 13 & 16\\
Snowden Defense Fund & 3 & 8 & 11\\
The Pirate Bay & 3 & 7 & 10\\
DarkWallet & 9 & 1 & 10\\
ProtonMail & 1 & 7 & 8\\
Darknet Mixer & 1 & 2 & 3\\
Liberty Hackers & 0 & 2 & 2\\
CryptoLocker Ransomware & 1 & 0 & 1\\
\bottomrule
\end{tabular}
}
\caption{Top-10 linked service providers.}
\label{table:linked_services}
\end{table}

Another way to present these results is from the standpoint of services. Table~\ref{table:linked_services} lists the top-10 service providers sorted by how many users were linked to them. The list is topped by WikiLeaks, which is a service that publishes secret information provided by anonymous sources, with 46 linked users. This is followed by Silk Road, the famous black market, with transactions from 22 users whose input coins have been seized by the FBI. Although the payment address of Silk Road was seized, it still appears in transactions until recently. However, based on further analysis, we found that a number of transactions were performed prior to the seizure. Ranked fifth, The Pirate Bay, which is known for infringing IP and copyright laws by facilitating the distribution of protected digital content, was linked to 10 users. As the linked users have accounts with various personally identifiable information (PII), they are vulnerable to the threat of deanonymizing their true identities. We next focus on two case studies that illustrate this threat in more detail.

\paragraph{Actionable links}
Purchasing products and services of black markets is generally considered illegal and calls for legal action. Some of the 22 users who are linked to Silk Road through transactions with seized coins shared enough PII to completely deanonymoize their identity. For example, one user is a 16 years old male from Crossville, Tennessee, U.S. The user has been a registered BitcoinTalk member since 2013, and has a transaction with Silk Road in October 2013, the takedown year, when he was around 13 years old. The corresponding user account points to his personal website, which contains links to his user profiles on Facebook, Twitter, and Youtube. Even if users do not share PII or use fake identities on their accounts, simply having an account on social networks is enough to track them online, or even secure a subpoena to collect identifiable information, such as login IP addresses. For example, three out of the 18 BitcoinTalk users recently logged in to the website.

\paragraph{A matter of jurisdiction}
One of the users who are linked to The Pirate Bay is a 36 years old male from Sweden. The Pirate Bay was founded by a Swedish organization called Piratbyr{\aa}n. Furthermore, the original founders of the website were found guilty in the Swedish court for copyright infringement activities. Since then, the website has been changing its domain constantly, and eventually operated as a Tor hidden service. Consequently, having such a link to The Pirate Bay through recent transactions in Sweden can lead to legal investigation, at least, and potentially be incriminating. 

\subsection{Market economics}
\label{sec:economics}

Keeping track of market statistics describing Tor hidden services is useful for identifying thriving services, measuring the impact of law enforcement, and prioritizing e-crime investigations. As such, an analyst may start with calculating a financial ``balance sheet'' for service providers, which typically includes the number of transactions with which a service is involved (i.e., volume), the amount of coins a service has received or sent (i.e., money flow), and the difference between the timestamps of the last and first transactions (i.e., operation lifetime). These statistics can be calculated in BlockTag using two queries, as shown in Listing~\ref{listing:market_stats}.

\begin{lstlisting}[caption={Calculating a balance sheet for service tags.},label={listing:market_stats}]
balance_sheet = chain.query(
    level='transaction',
    select= ['output.address.tag.info.provider as @name',
		'count(self.txes) as volume',
		'sum(input.value) as incoming',
		'min(time) as first_tx',
		'max(time) as last_tx',
		'date_diff(max(time), min(time)) as num_days'],
    where={
        'output': {
            'address': {
                'tag': {'type':'service', 'source':'tor'}
            }
        }
    },
    group_by='output.address.tag.info.id'
)
balance_sheet.join(
	results=chain.query(
		level='transaction',
		select= ['input.address.tag.info.provider as @name',
			'sum(input.value) as outgoing'],
		where={
			'input': {
				'address': {
					'tag': {'type':'service', 'source':'tor'}
				}
			}
		},
		group_by='input.address.tag.info.id'
	),
	on='@name' 
)
\end{lstlisting}

BlockTag supports joining queries via \texttt{results.join()} method of a query's results object. The join method operates on properties that can be aliased and referenced across queries using the \texttt{@alias} operator. In Listing~\ref{listing:market_stats}, the two queries are joined in order to calculate the money flow, as an address of a service tag can be an input or an output of a transaction, depending on whether it is an incoming or outgoing payment. Table~\ref{table:market_stats} shows the market statistics for the top-10 service providers ranked by incoming coins.

\begin{table*}
\centering
{\small
\begin{tabular}{l r r r l l r} \toprule
 & Volume & \multicolumn{2}{c}{Flow of money (\bitcoin)} & \multicolumn{3}{c}{Lifetime (dd/mm/yyyy)}\\
\cmidrule(lr){3-4}
\cmidrule(lr){5-7}
Name & (\# txs) & Incoming & Outgoing & First tx & Last tx & \# days\\ \midrule
Silk Road & 1,242 & 29,676.99 & 29,658.80 & 02/10/2013 & 19/03/2018 & 1,628\\
WikiLeaks & 26,399 & 4,043.00 & 4,040.74 & 15/06/2011 & 21/03/2018 & 2,470\\
VEscudero Escrow Service & 192 & 842.42 & 842.42 & 27/05/2012 & 20/08/2017 & 1,910\\
Internet Archives & 2,957 & 775.86 & 746.89 & 06/09/2013 & 21/03/2018 & 1,656\\
Freenet Project & 280 & 691.87 & 687.62 & 23/02/2011 & 16/03/2018 & 2,577\\
Snowden Defense Fund & 1,722 & 218.95 & 218.95 & 11/08/2013 & 18/03/2018 & 1,680\\
ProtonMail & 3,096 & 208.40 & 208.36 & 17/06/2014 & 18/03/2018 & 1,369\\ 
Ahmia Search Engine & 1,423 & 176.51 & 176.50 & 27/03/2013 & 06/03/2018 & 1,652\\
DarkWallet & 983 & 114.62 & 97.40 & 16/04/2014 & 02/11/2016 & 931\\
The Pirate Bay & 1,214 & 76.80 & 76.80 & 29/05/2013 & 21/08/2017 & 1,544\\
\bottomrule
\end{tabular}
}
\caption{Balance sheet of top-10 service providers ranked by incoming coins.}
\label{table:market_stats}
\end{table*}

\paragraph{Volume}
While the number of created service tags is small, the corresponding service providers have been involved in a relatively large number of transactions. For example, WikiLeaks tops the list with 26.4K transactions. The Darknet Mixer, which did not make it to the top-10 list in Table~\ref{table:market_stats}, has a volume of 22.1K transactions that is greater than the remaining services combined. One explanation for this popularity is that users are actually aware of the possibility of linking, and try to use mixing services in order to make traceability more difficult and improve their anonymity.

\paragraph{Money flow}
One interesting observation is that service providers have a nearly zero balance, which means almost the same amount of money comes in and goes out of their addresses. This indicates that the money is likely distributed to other addresses and is not kept on payment-receiving addresses. One explanation for this behavior is that by distributing funds among multiple addresses, a service provider can reduce coin traceability. Moreover, service providers still need to distribute their revenues among owners, sellers, and other stakeholders. Among all service providers listed in Table~\ref{table:market_stats}, Silk Road stands out with an income of $\bitcoin29.6$K.

\paragraph{Lifetime}
The services vary in their lifetime, ranging from two to seven years of operation. The first transaction date does not imply that the service provider began its operation on that date. It merely indicates the date on which the service provider started receiving payments through the tagged addresses. Looking at last transaction dates, all but three services are still active in 2018. For example, Silk Road has been receiving money since October 2013, even after the address has been seized by the FBI and its coins auctioned for sale by the U.S. Justice Department in June, 2014. However, a large number of post-seizure transactions appear to be novelty tips. 

\subsection{Forensics}
\label{sec:fraud_detection}

Organizations responsible for consumer protection, such as trade commission agencies and financial regulatory authorities, have a mandate to research and identify fraud cases involving cryptocurrencies, including unlawful initial coin offerings and Ponzi schemes. Given the popularity of Ponzi schemes in Bitcoin~\cite{vasek2015there,vasek2018analyzing}, we focus on this type of fraud and show how BlockTag can help analysts flag users who are likely victims or operators of such schemes.

A Ponzi scheme, also known as a high yield investment program, is a fraudulent financial activity promising unusually high returns on investment, and is named after a famous fraudster, Charles Ponzi, from the 1920s. The scheme is designed in such a way that only early investors will get benefits and once the sustainability of the scheme is at risk the majority of shareholders will lose the money they invested~\cite{artzrouni2009mathematics}. Among various Ponzi schemes in Bitcoin, MMM is considered one of the largest schemes that is hard to detect solely based on blockchain transaction analysis~\cite{bartoletti2018data}, highlighting the need for a systematic integration of auxiliary data into blockchain analysis. As such, an analyst can start the investigation with BlockTag using a full-text search query of keywords associated with MMM scheme, such as its name, without requiring prior knowledge of who is involved in the scheme or how it works, as shown in Listing~\ref{listing:text_search}.

\begin{lstlisting}[caption={Searching for tags using keywords.},label={listing:text_search}]
mmm_tags = chain.query(
	level='address',
	select= ['self.address', 'tag.id'],
	where={
		'tag': {
			'type': { '$in': ['user', 'text'] },
			'info': { '$like': 'mmm' }
		}
	}
)
\end{lstlisting}

This resulted in 24.2K user accounts, all of which are BitcoinTalk users, and 202 Blockchain.info text labels. For BitcoinTalk user accounts, the full-text search matched the \texttt{website} property of an account, which contained a URL pointing to the user's profile on MMM website. As for Blockchain.info text tags, the search matched the self-signed \texttt{label} property, which contained ``mmm'' substring, as summarized in Table~\ref{table:mmm_labels}. We next analyze the user accounts looking for clues related to MMM operation.

\paragraph{User demographics}
Out of 24.2K users, 52.86\%, 18.31\%, and 12.48\% shared their gender, age, and geo-location information, respectively. Based on this data, we found that the users are mostly male (75.44\%), between 20--40 years old (average=32), and are located worldwide in more than 80 different countries. However, 70.69\% of the users were located in only five countries, namely Indonesia, China, India, South Africa, and Thailand. Interestingly, most of these countries have a corresponding MMM label, as listed partially in Table~\ref{table:mmm_labels}.

\paragraph{Forum activity}
We used three properties of a user account that relate to activity on the forum, namely, \texttt{date\_registered}, \texttt{last\_seen}, and \texttt{num\_activities}. We found that 99.44\% of the users registered on the forum between August 2015--March 2016. Moreover, 98.21\% of the users made their last activity on the forum during the same period. This suggest that users have short-lived accounts. In fact, we found that 94.25\% of the users were active for 30 days or less, and that 78.45\% of users were dormant, meaning they were active for less than a day after registration. This also suggests that most of the users are not engaged with the forum. Indeed, only 313 users made at least one activity, and even for these users, they never engaged with the forum for more than once a day, on average. After manually inspecting the accounts on the website, we found that most of them were created as part of its ``MMM Extra'' scheme, which promises ``up to 100\% return per month for performing simple daily tasks that take 5--15 min,'' such as promoting MMM on social networks. This was evident from the accounts' signatures, which the crawler did not parse, that included messages such as ``MMM Extra is the right step towards the goal'' and ``MMM participants get up to 100\% per month.''

\paragraph{Financial operation}
As tags are linked through transactions in BlockTag, we can explore how MMM scheme operates financially through transaction graph analysis~\cite{ron2013quantitative}. In this analysis, Bitcoin transactions are modeled as a weighted, directed graph where nodes represent addresses, edges represent transactions, and weights represent information about transactions, such as input/output values and dates. Analyzing the topological properties of this graph can provide insights into which addresses are important and how the money flows. For example, having a few ``influential'' nodes and a small clustering coefficient suggest that most of the money funnels through these nodes and does not flow back to others, which are indicative of a Ponzi operation~\cite{vasek2015there,vasek2018analyzing,bartoletti2018data}. In BlockTag, an analyst can easily model case-specific transaction graphs by linking tags based on some search criteria, as shown in Listing~\ref{listing:linking_mmm}.

\begin{table}
\centering
{\small
\begin{tabular}{l r} \toprule
Label & Frequency\\ \midrule
mmm universe.help & 46\\
mmm global & 13\\
bonus from mmm universe.help & 9\\
mmm indonesia & 6\\
mmm nusantara & 4\\
mmm china & 2\\
mmm india & 2\\
mmm indonesia & 2\\
mmm philippines & 2\\
mmm russia & 2\\
\bottomrule
\end{tabular}
}
\caption{Top-10 frequent MMM labels.}
\label{table:mmm_labels}
\end{table}

\begin{lstlisting}[caption={Linking tags based on full-text search.},label={listing:linking_mmm}]
mmm_txes = chain.query(
	level='transaction',
	select= ['input.address.tag.id', 'output.address.tag.id', 'self.txes'],
	where={
		'input': {
			'address': {
				'tag': {
					'type': { '$in': ['user', 'text'] },
					'info': { '$like': 'mmm' }
				}
			}
		},
		'output': {
			'address': {
				'tag': {
					'type': { '$in': ['user', 'text'] },
					'info': { '$like': 'mmm' }
				}
			}
		}
	},
	group_by=['input.address.tag.type', 'output.address.tag.type'] 
)
\end{lstlisting}

\begin{figure}[!h]
\centering
\includegraphics[width=0.65\linewidth]{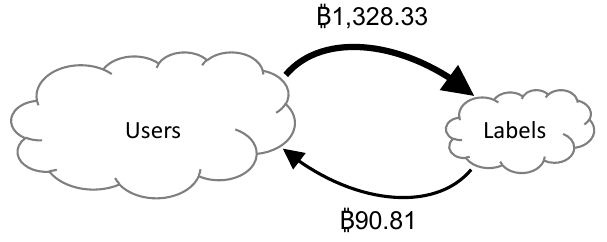}
\caption{MMM transaction graph.}
\label{fig:tx_graphs}
\end{figure}

We used the query in Listing~\ref{listing:linking_mmm} to model and analyze five transaction graphs, one for every combination of tag types, as summarized in Table~\ref{table:graph_stats}. The MMM transaction graph includes addresses of any type, and consisted of 14.3K addresses (i.e., order) and 32.5K transactions (i.e., size). This graph is also sparsely connected, as suggested by the small-sized largest strongly connected component (LSCC), low clustering, and long distance measures. Moreover, it consists of two subgraphs, the user$\rightarrow$user subgraph, which is also sparsely connected, and the label$\rightarrow$label subgraph, which is dense and small. Even though the two subgraphs are loosely connected through only 170 edges, an order of magnitude more money has flown from users to labels than the reverse direction, as shown Figure~\ref{fig:tx_graphs}.

\begin{table*}
\centering
{\small
\begin{tabular}{l l r r r r r r r r r r r} \toprule
 & & & & \multicolumn{4}{c}{Largest component} & \multicolumn{3}{c}{Clustering} & & \\
 \cmidrule(lr){5-8}
 \cmidrule(lr){9-11}
\multicolumn{2}{c}{Type} & & & \multicolumn{2}{c}{Weakly connected} & \multicolumn{2}{c}{Strongly connected} & & \multicolumn{2}{c}{Triangles} & \multicolumn{2}{c}{Distance (LSCC)}\\
 \cmidrule(lr){1-2}
 \cmidrule(lr){5-6} 
 \cmidrule(lr){7-8}
 \cmidrule(lr){10-11}
  \cmidrule(lr){12-13}
Input & Output & Order & Size & \#nodes & \#edges & \#nodes & \#edges & Average & \#triangles & \%closed & Diameter & Radius\\ \midrule
User & User & 14,227 & 31,819 & 13,914 & 31,631 & 5,850 & 17,498 & 0.11 & 6,566 & 0.08 & 17 & 7\\
User & Label & 129 & 125 & 96 & 103 & 1 & 0 & 0.00 & 0 & 0.00 & 0 & 0\\
Label & User & 64 & 45 & 10 & 9 & 1 & 0 & 0.00 & 0 & 0.00 & 0 & 0\\
Label & Label & 61 & 508 & 54 & 498 & 20 & 246 & 0.64 & 943 & 61.04 & 3 & 2\\
Any & Any & 14,319 & 32,497 & 14,002 & 32,307 & 5,934 & 18,128 & 0.11 & 7,576 & 0.09 & 17 & 7\\
\bottomrule
\end{tabular}
}
\caption{Properties of MMM transaction (sub)graphs.}
\label{table:graph_stats}
\end{table*}

To find influential nodes in the graph, we computed their PageRank, where weights represented input address values of transactions. All of the top-10 ranked nodes were located in the user$\rightarrow$user subgraph, which mapped to unique BitcoinTalk users. After manually inspecting the corresponding accounts, we found that the first and the third users have been reported as scammers on BitcoinTalk for operating fraudulent services, namely Dr.BTC and OreMine.Org. While the first user has received a total of $\bitcoin$426.7K on her address, the third has received a staggering total of $\bitcoin$1.8M on his address that is associated with Huobi wallet address, an exchange service, suggesting that the user has exchanged the received coins.

\section{Discussion}
\label{sec:discussion}

In what follows, we discuss the limitations of our work and outline our plan for current and future work.

\paragraph{Limitations}

BlockTag's main limitation is the validity of its tags, since they are created automatically by crawlers from open, public data sources. This limitation is part of a larger problem that is common with Internet content providers, such as Google and Facebook, especially when this content is generated mostly by users~\cite{yin2008truth,li2016survey}. In general, the validity issue is especially important for user identities, as attackers and fraudsters can always create fake accounts in order to hide their real identity~\cite{ferrara2016rise}. While doing so improves their anonymity, law enforcement agencies can use the links found through BlockTag to secure a subpoena in order to collect more information about suspects from website operators~\cite{marquardt2014dpr}.

\paragraph{Work in progress}
We are designing BlockSearch, an open-source Google-like searching layer that sits on top of BlockTag. BlockSearch allows analysts to search blockchain for useful information in plain English and in real-time, without having to go through the hassle of performing low-level queries using BlockTag. The system also provides in a dashboard for analysts that displays real-time results of important queries, such as the ones we presented in the paper. Based on feedback from trade commission agencies and financial regulatory authorities, such capabilities are extremely helpful to protect customers, comply with know you customer (KYC) and anti-money laundering (AML) laws, and draft new, investor-friendly cryptocurrency regulations.

\paragraph{Future work}
In order to address the main limitation of BlockTag, we plan to define confidence scores for tag sources. The scores can be computed using various ``truth discovery'' algorithms~\cite{dong2009integrating}, which are generally based on the intuition that the more sources confirm a tag the more confidence is assigned to it.

BlockTag is modular by design. This means we can easily enhance or add new capabilities. As such, we plan to implement more vertical crawlers for services such as WalletExplorer,\footnote{\url{https://www.walletexplorer.com}} ChainAlysis,\footnote{\url{https://www.chainalysis.co}} BitcoinWhosWho,\footnote{https://bitcoinwhoswho.com} and Reddit.\footnote{\url{https://www.reddit.com}} We also plan to support more clustering methods and develop a systematic way to automatically tag clusters, in addition to blocks, transactions, and addresses, based on label propagation algorithms~\cite{gregory2010finding}.

\section{Conclusion}
\label{sec:conclusion}

Blockchain analysis has become a hot topic among researchers and law enforcement agencies for applications that demand more effective tools. While state-of-the-art analysis systems, such as BlockSci, are efficient, they are not designed to annotate and analyze auxiliary blockchain data. We presented BlockTag, an open-source tagging system for blockchains. We used BlockTag to uncover privacy issues with using Bitcoin in Tor hidden services, and flag Bitcoin addresses that are likely to be part of a large Ponzi scheme.

\section*{Acknowledgements}

We would like to thank the people at the Cybersecurity Initiative for Blockchain Research (CIBR) for their help and feedback.\footnote{For latest research outcomes, please visit \url{https://qcri.github.io/cibr}} This research was partially supported by Qatar National Research Fund through grant \#QNRF-AICC01-1228-170004.

\bibliographystyle{ACM-Reference-Format}
\bibliography{refs}

\end{document}